\documentstyle[12pt]{article}
 
\newcommand{\be}{\small\begin{equation}}
\newcommand{\ee}{\end{equation}\normalsize\vspace*{-0.1ex}}
\newcommand{\bea}{\small\begin{eqnarray}}

\newcommand{\eea}{\end{eqnarray}\normalsize\vspace*{-0.1ex}}
\newcommand{\bdm}{\small\begin{displaymath}}
\newcommand{\edm}{\end{displaymath}\normalsize\vspace*{-0.1ex}}
\newcommand{\beas}{\small\begin{eqnarray*}}

\newcommand{\eeas}{\end{eqnarray*}\normalsize\vspace*{-0.1ex}}
  
\begin{document}
 
 
\thispagestyle{empty}
\renewcommand{\thefootnote}{\fnsymbol{footnote}}
 
\setcounter{page}{0}
\begin{flushright} MPI-PhT/94-18\\
UM-TH-94-13\\
May 1994  \end{flushright}
 
\begin{center}
\vspace*{2cm}
{\Large\bf
Bloch-Nordsieck
cancellations beyond logarithms in heavy particle decays
}\\
\vspace{1.8cm}
{\sc M.Beneke}${}^1$, {\sc V.M.Braun}${}^{2,}$\footnote{
On leave of absence from
St.Petersburg Nuclear Physics Institute, 188350 Gatchina,
Russia} and
{\sc V.I.Zakharov}${}^{1,}$\footnote{
On leave of absence from
Max-Planck-Institut f\"ur Physik, F\"ohringer Ring 6, 80805
Munich}\\[0.5cm]
\vspace*{0.3cm} ${}^1$ {\it Randall Laboratory of Physics\\
University of Michigan\\ Ann Arbor, Michigan 48109, U.S.A.}\\[0.6cm]
\vspace*{0.3cm} ${}^2${\it Max-Planck-Institut
f\"ur Physik\\ -- Werner-Heisenberg-Institut -- \\
F\"ohringer Ring 6\\ D--80805
Munich (Fed. Rep. Germany)}\\[1.4cm]
 
{\bf Abstract}\\[0.3cm]
\end{center}
 
We investigate the one-loop radiative corrections to the semileptonic
decay of a charged particle at finite gauge boson mass. Extending the
Bloch-Nordsieck cancellation of infrared  logarithms, the
subsequent non-analytic terms are also found to vanish after
eliminating the pole mass in favor of a mass defined at short
distances. This observation
justifies the operator product expansion for
inclusive decays of heavy mesons 
and implies that infrared effects associated with
the summation of the radiative corrections are suppressed by at
least three powers of the mass of the heavy decaying particle.\\
 

\newpage
\renewcommand{\thefootnote}{\arabic{footnote}}
\setcounter{footnote}{0}
 
 
The problem of the infrared (IR) behavior of amplitudes is inherent to
gauge theories since by construction they contain massless bosons.
 Historically, already the first studies of the problem revealed both
 complexities and simplicities. On the one hand, if a finite gauge boson
mass $\lambda$ is employed as IR regulator, the perturbative amplitudes
contain logarithms in the mass, $\ln\lambda^2$, which apparently do not
allow for the limit $\lambda\rightarrow 0$. On the other hand, as shown
first by Bloch and Nordsieck \cite{BLO37}, these singularities cancel if one
 considers inclusive processes. This summation over final states
is an integral part of any calculation in gauge theories.
 
With the advent of QCD the problem of the IR behavior
of amplitudes became even more acute since the effective coupling blows up
in the infrared. Thus, if an amplitude is perturbatively sensitive to infrared
momenta, it is contributed in fact by nonperturbative effects as well
and cannot be evaluated 
reliably. From this point of
view the famous prediction of QCD that the total cross section
of $e^+e^-$ into hadrons measures quark charges rests on the Bloch-Nordsieck
(BN) cancellation. Indeed, all the IR
logarithms disappear from the total cross section and it is
determined by short distance physics. Technically, $e^+ e^-$
annihilation is treated within
the operator product expansion (OPE). Then, at least naively, IR
contributions are suppressed by four powers of the large scale.
The BN type cancellations ensure that this counting is not violated
by divergencies of perturbative expansions, explicit or
through its divergence in large orders. In this sense there
is a correspondence between IR cancellations and the OPE.
 
The example of
$e^+e^-$ annihilation into hadrons
is still special since there are no colored particles
in the initial state. In the general situation, it is known
from the work of Kinoshita and Lee and Nauenberg (KLN) \cite{KIN62}
that to remove all IR singularities, summation
over degenerate initial states might also be needed.
This summation does not necessarily reflect the
actual experimental situation and in this sense some IR logarithms can
survive.
 
The OPE has been applied more recently in
the context of heavy quark physics and the inclusive decay widths
of heavy hadrons in particular. In this case there is a charged particle in
the initial state. Though it is generally accepted that with
only one colored particle in the initial state, the BN theorem
is sufficient to ensure the absence of {\it explicit} soft divergencies,
it is not a priori clear whether
one can rely on {\it extended} BN type cancellations in the sense of
IR power counting, implying the validity of OPE,
or whether one has to invoke the more general KLN cancellation,
which would require modifications of the
standard OPE. In particular, the standard OPE states that there are no
corrections
to the decay width linear in the inverse mass of the heavy quark.
This conclusion is not trivial
in view of the recent observation \cite{BIG94,BEN94} that
the pole mass of a charged particle itself does receive such corrections.
It has been argued, therefore, that the
widths are to be expressed in terms of a (unphysical) quark mass
normalized far off-mass-shell,
which is infrared stable in the linear approximation \cite{BIG94}.

In this Letter we test the counting of IR effects implied by
the OPE by an explicit
calculation of the IR sensitive contributions to the decay of a 
charged particle.
The infrared contributions are singled out by non-analytic terms 
in the gauge
boson mass, some of which are associated with higher dimensional 
operators. The ones which are not should disappear either after 
summation over final states or final and initial states. In the 
latter case the IR counting implied by the OPE is violated. 
The leading  $\ln \lambda^2$ terms are the subject
of the original BN cancellation \cite{BLO37}, and we show that the
BN mechanism extends
to comprise the $\sqrt{\lambda^2}$ and $\lambda^2\ln\lambda^2$ terms,
if the physical (or pole)
mass is eliminated in favor of a short distance
mass.
The cancellation of linear in $\lambda$ terms is in accord with
the arguments of Ref. \cite{BIG94}.
Note that although we use the QCD terminology, present calculations are 
not sensitive to the non-abelian nature of QCD and our
results are strictly speaking applicable to the abelian case only.

For definiteness, we consider the decay of a $B$ meson.
To lowest order in the weak coupling,
the QCD dynamics is contained in the matrix element of the forward
scattering operator of the product of two weak hadronic currents
between $B$-meson states
 
\bdm
T^{\mu\nu}(p,q) = i\int\mbox{d}^4 x\,e^{i q x}\langle B(p)|
T\left\{{J_h^\mu}^\dagger (x) J_h^\nu (0)\right\}|B(p)\rangle ,\edm
 
\noindent where an average over initial state polarizations is 
understood. The differential inclusive width is then given by
 
\be\label{diffwidth}
\mbox{d}\Gamma = \frac{G_F^2 |V_{qb}|^2}{2}
\frac{\mbox{d}^3 k_{\bar{\nu}}}{(2\pi)^3 2 k^0_{\bar{\nu}}}
\frac{\mbox{d}^3 k_l}{(2\pi)^3 2 k^0_l}
\,L_{\mu\nu}\cdot 2\,\mbox{Im}\left[\frac{1}{2 p^0}
T^{\mu\nu}(p,q)\right] .\ee
 
\noindent The lepton tensor $L_{\mu\nu}$ contains the lepton 
momenta
$k_l$, $k_{\bar{\nu}}$ only and the imaginary part is taken in
$p\cdot q$, where $q=k_l+k_{\bar{\nu}}$.
Assuming validity of the OPE a systematic expansion in
powers of the heavy quark mass can be performed for the lepton
spectrum and results in the prediction \cite{BIG92}
 
\be\label{totalwidth}
\Gamma_B\,=\,\Gamma_0\left[1+\sum_{n=0}^\infty r_n\alpha(m_b)^{n+1}
+ \frac{\mu_K-3\mu_G}{2 m_b^2} + O\left(\frac{1}{m_b^3}\right)
\right]\ee
 
\noindent for the total width. Here $m_b$ is the ${\it pole}$ quark 
mass, $\Gamma_0 =
(G_F^2|V_{qb}|^2 m_b^5)/(192\pi^3)$ the tree decay width for the
free quark and the first order radiative correction is known
explicitly \cite{BEH56}.
For simplicity, the final quark $q$ is taken
massless.
 
A striking feature
of Eq. (\ref{totalwidth}) is that corrections to the free quark
decay are suppressed by {\it two} powers of the quark mass. They
can be parametrized by
 
\be\label{matrixelements}
\mu_K = \frac{1}{2 m_B}\langle B|\bar{b}(iD_\perp)^2 b|B\rangle
\qquad \mu_G = \frac{g}{4 m_B}\langle B|\bar{b} i\sigma_{\mu\nu}
G^{\mu\nu} b|B\rangle ,\ee
 
\noindent where $m_B$ is the meson mass and we ignore the radiative
corrections to these terms.
The crucial assumption to arrive at this conclusion is the OPE
in the kinematic region where the energy release into the hadronic 
final state is large and the decaying quark is almost on-shell. 
In addition, 
radiative corrections to the leading operator
must be unambiguous to this accuracy.
The near-mass-shell condition
for the decaying quark is a new
ingredient compared to the familiar expansions of
the same current product on the light cone or at short distances. 
In particular,
the pole mass which provides the overall normalization in Eq. (3)
is intrinsically ambiguous by an amount of order
$\Lambda_{QCD}$ \cite{BIG94,BEN94}.
The perturbative series
that relates the pole mass to a mass defined at short distances,
e.g. the $\overline{MS}$-mass,
 
\be \label{massseries}
m_b = m_b^{\overline{MS}}\left(1+\sum_{n=0}^\infty c_n
\alpha^{n+1}\right) ,\ee
 
\noindent exhibits a strong divergence as $n$ increases, which leads
to an IR renormalon pole in its Borel transform at $t=-1/(2\beta_0)$
with $t$ the Borel parameter (to be defined below) and $\beta_0$
the first coefficient of the $\beta$-function. Truncating the series
at its optimal order leaves an uncertainty of order $\Lambda_{QCD}$,
assuming $m_b^{\overline{MS}}$ as given.
 
The important question is whether the heavy quark expansion
which relies on the heavy quark being near mass-shell -- a notoriously
IR-singular point -- captures {\it all} IR effects {\it after} the
effects associated with the definition of mass have been accounted for.
To clarify this point, we have investigated the IR structure of the
asymptotic behavior of the radiative corrections to the leading term
in Eq. (\ref{totalwidth}). Through the
appearance of IR renormalons the asymptotic behavior signals the presence
of power corrections, which should be added as explicit nonperturbative
corrections. These explicit corrections may -- and often do -- turn
out numerically larger than higher order perturbative corrections,
which are neglected. Any indication of such terms of order $1/m_b$
threatens the validity of the OPE for inclusive decays.
 
We emphasize that the absence of IR renormalons can be stated
as an extension of the BN cancellations.
Such a relation might be envisaged, since the IR renormalons
probe the IR behavior of Feynman amplitudes and can therefore
be traced by an IR regulator.
To establish a formal connection, we recall that the large-order
behavior of radiative corrections to first approximation is generated by
diagrams such as in Fig. 1, with a chain of fermion bubbles inserted
in the gluon line.
This procedure generates a
gauge-invariant set of diagrams in the
abelian theory. Denote by $\{r_n^f\}$ the series of
perturbative corrections to $B$-decays ($\mu^+ e^-$-decays)
generated in this way, and define the Borel transform of the
series by $B[\{r_n^f\}](t)\equiv\sum_{n=0}^\infty r_n^f\,
t^n/n!\,$. The fermion loop insertions are renormalized
and each loop is proportional to $(-\beta_0)[\ln(-k^2/\mu^2)
+C]$, where $\beta_0$ includes the fermionic contribution only,
$\mu$ is a renormalization point and $C$ a finite subtraction
constant. Next call $r_0(\lambda)$ the one-loop radiative
correction calculated with a finite gluon (photon) mass. To
ensure the existence of the zero mass limit, we keep the
standard gauge fixing and work with the propagator
 
\be\label{gluonprop}
-i\delta^{AB}\frac{1}{k^2-\lambda^2+i\epsilon}
\left[g_{\mu\nu} - (1-\xi) \frac{k_\mu k_\nu}{k^2-\xi\lambda^2
+i\epsilon}\right] .\ee
 
\noindent Then, in the Landau gauge, $\xi=0$, one finds the 
identity
 
\be\label{relation}
r_0(\lambda)\,=\,\frac{1}{2\pi i}
\!\!\!\int\limits_{-1/2-i\infty}^{-1/2+
i\infty}\!\!\!\!\!\mbox{d} s \,\Gamma(-s)\Gamma(1+s)\left(\frac{
\lambda^2}{\mu^2} e^C\right)^s\,B[\{r_n^f\}](s)
\ee
 
\noindent with $s=-\beta_0 t$. To obtain this identity, one uses that
the effective Borel-transformed propagator of the massless gauge boson
is proportional to $1/(k^2)^{1+s}$ after summation over the fermion
loops \cite{BEN93}, whereas the propagator,
Eq. (\ref{gluonprop}), can be written in a Mellin representation,
 
\bdm
\frac{1}{k^2-\lambda^2} = \frac{1}{2\pi i}\,\frac{1}{k^2}
\!\!\!\int\limits_{-1/2-i\infty}^{-1/2+i\infty}
\!\!\!\!\!\mbox{d} s\,\Gamma(-s)
\Gamma(1+s)\left(-\frac{\lambda^2}{k^2}\right)^s .\edm
 
\noindent The significance of Eq. (\ref{relation}) rests on the
observation that the coefficients in front of the
$(\sqrt{\lambda^2})^{2 n+1}$ and $\lambda^{2 n}\ln\lambda^2$ terms
of the expansion of $r_0(\lambda)$ in the small mass determine
the residues of the IR renormalon poles of $B[\{r_n^f\}](t)$
at half-integer and integer
multiples of $-1/\beta_0$, which in turn fixes the overall
normalization of the large-order behavior of the series
${r_n^f}$. In particular, cancellations of terms in the
expansion of $r_0(\lambda)$ imply the absence of the
corresponding renormalons. In this unified framework, the
BN cancellations of $\ln \lambda^2$ in physical processes
appear simply as the absence of an IR renormalon pole
at $s=0$ \cite{remark1}.
Note that the analytic terms $\lambda^{2n}$ in the
expansion of $r_0(\lambda)$ are of purely kinematic origin and
not related to renormalons. Since the Borel transform
is gauge-independent for physical processes and the $S$-matrix
elements of the abelian gauge theory are independent of $\xi$
even in the presence of a mass term, it follows, that Eq.
(\ref{relation}) holds in fact in any gauge.
 
Using the relation between IR renormalons and non-analytic in
$\lambda^2$ terms we can rewrite the relation between
the pole and a short-distance mass as
 
\be\label{sdmass} m_b \sim m_{SD}\, (1-2\tilde{\lambda})\qquad
\quad\tilde{\lambda}\equiv C_F\alpha/(4\pi)\times
\pi\lambda/m_b\,.\ee
 
\noindent The presence of linear terms in this place is equivalent 
to the asymptotic behavior $c_n\stackrel{n\rightarrow \infty}{=}
(C_F\mu e^{-5/6})/(\pi m_b^{\overline{MS}}) (-2\beta_0)^n n!$
for the series in Eq. (\ref{massseries})
obtained in \cite{BIG94,BEN94}, which gives rise to the
uncertainty of order $\Lambda_{QCD}$. For our purposes here we
do not need to specify $m_{SD}$
and a renormalization point any further.
 
To check the infrared effects in the decay widths
we have calculated the radiative corrections
to the total width and the lepton spectrum
keeping the gluon mass finite (and the final quark massless).
The Mellin representation
of the gluon propagator is useful technically, since
often it turns out to be easier to obtain the relevant
terms in the expansion in $\lambda$ by closing the contour in the
right plane and picking out the relevant residues rather than to
calculate the integrals exactly and then expand in the mass. The
resulting expressions for the diagrams of Fig. 1
in $4-2\epsilon$ dimensions are (``$\sim$''
means: the $1/\epsilon$, $\ln\lambda^2$, $\sqrt{\lambda^2}$
and $\lambda^2\ln \lambda^2$
terms of the l.h.s. equal the r.h.s.):
 
\bdm T^{\mu\nu}_{(b)}(p,q) \sim T^{\mu\nu}_{(a)}(p,q)\,\frac{C_F\alpha}
{4\pi}\left[-\frac{\xi}{\epsilon}-(3-\xi)\ln\frac{\lambda^2}{m_b^2}+
3\pi\frac{\lambda}{m_b}\right]
\edm
\bdm
T^{\mu\nu}_{(c)}(p,q) \sim T^{\mu\nu}_{(a)}(p,q)\,\frac{C_F\alpha}
{4\pi}\left[-\frac{\xi}{\epsilon} +
\left(2-\frac{1-\xi^2}{2}\right)
\frac{\lambda^2}{p_q^2+i\epsilon}
\ln\left(-\frac{\lambda^2}{p_q^2}\right)
\right]
\edm
\bea\label{diagrams}
T^{\mu\nu}_{(d)}(p,q) &\sim& T^{\mu\nu}_{(a)}(p,q)\,\frac{C_F\alpha}
{4\pi}\left[\frac{2 \xi}{\epsilon}-4\pi\omega
\frac{\lambda}{m_b}\right] -\frac{1}{p_q^2+i\epsilon}\,
\mbox{tr}\left(\not\!p\Gamma^\mu\not\!p\Gamma^\nu\right)\,\frac{C_F\alpha}
{4\pi}\left[2\pi\frac{\lambda}{m_b}\right]
\nonumber\\
&& +\,T^{\mu\nu}_{(a)}(p,q)\,\frac{C_F\alpha}
{4\pi}\left[1-\xi^2\right]\frac{\lambda^2}{p_q^2+i\epsilon}
\ln\left(-\frac{\lambda^2}{p_q^2}\right)\nonumber\\
&& -
\frac{1}{p_q^2+i\epsilon}\,
\frac{C_F\alpha}
{4\pi}\Bigg[
 \frac{ m_b^2}{p_q^2+i\epsilon}\,
\mbox{tr}\left(\not\!p_q\Gamma^\mu\not\!p_q\Gamma^\nu\right)
 -\frac{1}{2} m_b^2\,\mbox{tr}\left(\gamma_\tau\Gamma^\mu
\gamma^\tau\Gamma^\nu\right)
\nonumber\\ && +
\Bigg(\omega^2-2\omega-\frac{m_b^2}{p_q^2+i\epsilon}\,\Bigg)
\mbox{tr}\left(\not\!p\Gamma^\mu\not\!p_q\Gamma^\nu\right)
 -(\omega-2)\,
\mbox{tr}\left(\not\!p\Gamma^\mu\not\!p\Gamma^\nu\right)
\Bigg]\,\frac{\lambda^2}{m_b^2}\,
\ln\frac{\lambda^2}{m_b^2}\nonumber
\eea
\bea T^{\mu\nu}_{(e)}(p,q) &\sim& T^{\mu\nu}_{(a)}(p,q)\,\frac{C_F\alpha}
{4\pi}\left[(3-\xi)\ln\frac{\lambda^2}{m_b^2}+
\pi(2\omega-3)\frac{\lambda}{m_b}\right]\\
&& - \frac{1}{p_q^2+i\epsilon}\,
\mbox{tr}\left(\not\!p\Gamma^\mu\not\!p\Gamma^\nu\right)\,\frac{C_F\alpha}
{4\pi}\left[-\pi\frac{\lambda}{m_b}\right]
\nonumber\\
&& +\,T^{\mu\nu}_{(a)}(p,q)\,\frac{C_F\alpha}
{4\pi}\left[-\frac{1-\xi^2}{2}\right]\frac{\lambda^2}{p_q^2+i\epsilon}
\ln\left(-\frac{\lambda^2}{p_q^2}\right)
\nonumber\\
&& -
\frac{1}{p_q^2+i\epsilon}\,
\frac{C_F\alpha}
{4\pi}\Bigg[
- \frac{ m_b^2}{p_q^2+i\epsilon}\,
\mbox{tr}\left(\not\!p_q\Gamma^\mu\not\!p_q\Gamma^\nu\right)
 +\frac{1}{2} m_b^2\,\mbox{tr}\left(\gamma_\tau\Gamma^\mu
\gamma^\tau\Gamma^\nu\right)
\nonumber\\&&
-(\omega^2-2\omega)\,
\mbox{tr}\left(\not\!p\Gamma^\mu\not\!p_q\Gamma^\nu\right)
 +(\omega-2)\,
\mbox{tr}\left(\not\!p\Gamma^\mu\not\!p\Gamma^\nu\right)
\Bigg]\,\frac{\lambda^2}{m_b^2}\,
\ln\frac{\lambda^2}{m_b^2}\nonumber
\eea
 
\noindent Here $\Gamma^\mu=\gamma_\mu (1-\gamma_5)$, $C_F=4/3$,
$p_q=p-q$, $\omega = 2(p\cdot p_q)/(p_q^2+i\epsilon)$ and the tree
diagram is given by
 
\bdm T_{(a)}^{\mu\nu}(p,q) = -\frac{1}{2}\frac{1}{p_q^2+i\epsilon}\,
\mbox{tr}\left(\not\!p\Gamma^\mu\not\!p_q\Gamma^\nu\right) .\edm
 
\noindent Diagram (b) is to be understood as an on-shell wavefunction
renormalization and $m_b$ is the pole mass.
The ultraviolet (UV) divergent terms vanish in the sum of all
diagrams as well
as the logarithms in $\lambda$ as implied by the BN theorem.
In addition, the $\lambda^2\ln\lambda^2$ terms add to zero.
 
Note that the terms linear in
$\lambda$ do not yet cancel. The reason is the use of the pole mass
which contains terms linear in $\lambda$, see Eq. (\ref{sdmass}).
One may now eliminate
the pole mass from the differential width with the result
($p_{SD}^2=m_{SD}^2$)
 
\bdm\frac{1}{2 p^0}\,T^{\mu\nu}(p,q)\sim
\frac{1}{2 p_{SD}^0}\left(-\frac{1}{2}\right)
\frac{1}{p_{q,SD}^2+i\epsilon}\,
\mbox{tr}\left(\not\!p_{SD}\Gamma^\mu\not\!p_{q,SD}\Gamma^\nu
\right)\edm
 
\noindent with {\it no} corrections linear in $\lambda$ 
to first order
in $\alpha$. Since all dependence on $m_b$ has been eliminated
in Eq. (\ref{diffwidth}), it is never reintroduced through the
subsequent phase space integrations \cite{remark2}. Integrating
the differential width with the radiative corrections, Eq.
(\ref{diagrams}), which has to be done with care for a massless
final quark, gives
 
\be \label{width}
\Gamma_B\sim \frac{G_F^2 |V_{qb}|^2 m_b^5}{192 \pi^3}\,
\left(1 + \tilde{\lambda}\{3_b+20_d-13_e\}\right) .\ee
 
\noindent The subscript indicates the diagram, 
from which the respective
terms originate. This form makes the cancellation of linear terms,
when $m_b$ is replaced through Eq. (\ref{sdmass}),
more explicit. Let us rephrase the meaning of Eq. (\ref{width})
in the language of renormalons: The presence of linear terms
in the gluon mass implies an IR renormalon at $t=-1/(2\beta_0)$
in the asymptotic behavior of the radiative corrections to the
free quark decay in Eq. (\ref{totalwidth}). However, this divergent
behavior is organized in such a way, that it exactly cancels
against the IR renormalon in the series, Eq. (\ref{massseries}),
when the pole mass in the overall normalization of the width
is replaced by a short-distance mass. This cancellation supports
the validity of the OPE in the kinematic situation specified by
a decaying hadron. Even though the initial state is non-trivial,
to eliminate the non-analyticity one does not have to invoke a
summation over degenerate initial states, which would go beyond
an OPE treatment. Note that the disappearance of linear terms
in the total width has
already been concluded in Ref. \cite{BIG94} on basis 
of the applicability of the OPE \cite{exp}.

As already stated above, the subsequent non-analytic term 
$\lambda^2\ln\lambda^2$,
which is related to an IR renormalon at $t=-1/\beta_0$, adds
to zero in the sum of all contributions in Eq. (\ref{diagrams}).
Note that this non-analyticity is not reintroduced when
the pole mass is eliminated, since the mass shift, 
Eq. (\ref{sdmass}), does
not contain such terms, when $m_{SD}$ is a $MS$-like mass.
Thus, IR effects
associated with the
summation of the radiative corrections to the free quark decay are 
suppressed by {\it three} powers of the heavy quark mass.
This may be anticipated, since explicit nonperturbative corrections 
of order $1/m_b^2$ are present only due the spin interaction and the 
kinetic energy of the heavy quark (cf. Eq. (\ref{matrixelements})), 
both of which are zero for the free quark decay. The  
matrix elements between $B$ states are not affected by an 
IR region in the Feynman diagrams for the free quark decay. 
This is clear for $\mu_G$, the spin energy, which can 
be related to an observable, the mass difference of the vector and 
pseudoscalar mesons, to leading order, but less clear for 
the kinetic energy contribution, given the problematic 
aspects of the definition of a physical mass. 
Vanishing of the kinetic energy for a charged particle is 
a consequence of Lorentz symmetry, 
and is related to the absence of an IR renormalon in 
the pole mass at $t=-1/\beta_0$, observed in Ref. \cite{BEN94}. 
Combining this observation with the heavy quark expansion for the 
meson masses, it follows that the kinetic energy matrix element 
between $B$ states is protected form ambiguities due to 
renormalons.

We conclude that the suppression of infrared effects implied by
the OPE is indeed valid
for heavy particle decays.
Phenomenologically, the subtleties associated with the definition of
mass are only of secondary importance, as long as one does not attempt
a determination of the (unphysical) mass parameter. One may always
sacrifice one measurement to predict another quantity and use any
mass parameter in intermediate steps, in particular the pole mass, which
has convenient properties within perturbation theory in low orders.
The prime question
concerns the IR properties of the widths, after the IR properties of
the mass parameter have been abstracted. The presented cancellation is
an important step in proving that inclusive decay rates
of heavy quarks are infrared stable (in an extended sense)
and can be reliably calculated within
perturbation theory. However, our result applies strictly speaking to
the abelian case. It is not obvious that degenerate initial states
are equally unimportant in the non-abelian case as they are in
the abelian.
To prove this
is a challenge
yet to be met.\\
 
We are grateful to A.Vainshtein for an interesting
discussion and valuable comments.
M.B. is supported by the Alexander von Humboldt Foundation.
 
 

 

\begin{thebibliography}{99}
\bibitem{remark1} Some care is required to disentangle the
IR pole at $s=0$ from an UV pole at the same position, related
to renormalization.
\bibitem{remark2} A subtle point is that, when taking the
imaginary part in $p\cdot q$, one has to insert $\Theta(p^0-q^0)$
according to the Cutkosky rules. Shifting the mass according to
Eq. (\ref{sdmass}) produces a term proportional to
$\tilde{\lambda}\,\delta(p^0_{SD}-q^0)$, which is not compensated
by the radiative corrections, Eq. (\ref{diagrams}), but does not
contribute to the lepton spectrum and the total width.
\end{thebibliography}
\end{document}